\documentclass{webofc}
\usepackage{svg}
\usepackage[varg]{txfonts}   
\usepackage{hyperref}
\usepackage{url}
\hypersetup{colorlinks=true,citecolor=blue,urlcolor=blue,linkcolor=blue}
\begin{document}
\title{Fluid dynamics of
charm quarks\\
from heavy to light-ion collisions}
%
%

\author{
\firstname{Federica} \lastname{Capellino}\inst{1}\fnsep\thanks{\email{f.capellino@gsi.de}}
\and
\firstname{Andrea} \lastname{Dubla}\inst{1}
\and
\firstname{Rossana} \lastname{Facen}\inst{2}
\and \firstname{Stefan} \lastname{Floerchinger}\inst{3}
\and \firstname{Eduardo} \lastname{Grossi}\inst{4}
\and \firstname{Andreas} \lastname{Kirchner}\inst{5}
\and
\firstname{Silvia} \lastname{Masciocchi}\inst{1,2}
}
\institute{
GSI Helmholtzzentrum für Schwerionenforschung, 64291 Darmstadt, Germany \and
Universität Heidelberg, Physikalisches Institut, 69120 Heidelberg \and
Friedrich-Schiller-Universität Jena, 07743 Jena, Germany\and
Università di Firenze and INFN Sezione di Firenze,  50019 Sesto Fiorentino, Italy \and
Department of Physics, Duke University, Durham, NC 27708, USA 
}

\abstract{Heavy quarks are powerful tools to characterize the quark-gluon plasma (QGP) produced in relativistic nuclear collisions. By exploiting a mapping between transport theory and hydrodynamics, we developed a fluid-dynamic description of heavy-quark diffusion in the QCD plasma. We present results for the transverse momentum distributions of charm hadrons and evolution of charm density and diffusion fields obtained using a fluid-dynamic code coupled with the conservation of a heavy-quark current in the QGP in various collision systems. 
}
\maketitle
\section{Introduction}
\label{intro}
Charm quarks are produced in relativistic nuclear collisions via hard partonic scattering processes. Due to their large mass and early production, they are suitable probes for studying the QGP. Recent experimental measurements~\cite{ALICE:2020iug, ALICE:2020pvw} showed that J/$\psi$ and D mesons display a positive elliptic flow, suggesting thermalization of charm quarks within the QGP. The idea of charm thermalization is at the same time supported by theoretical and phenomenological studies (see e.g. ~\cite{Andronic:2021erx, Altenkort:2023oms}), which suggest that charm quarks have a short relaxation time. 

In our previous work~\cite{Capellino:2022nvf}, the question of charm thermalization was addressed by studying the time needed by charm quarks to fall into a hydrodynamic regime (\textit{hydrodynamization}) in the context of an expanding medium. It was shown that the time required for charm quark to hydrodynamize, and therefore to be included in the fluid-dynamic description of the QGP, is shorter than the typical expansion time scale of the plasma. This result served as motivation to develop a fluid-dynamic description of charm quarks, which turned out to be successful in describing the low-transverse momentum region of charm-hadron spectra as well as their integrated yields in Pb-Pb collisions at the LHC at $\sqrt{s_{NN}}=5.02$ TeV~\cite{Capellino:2023cxe}.

In this study, we extend our approach to charm quark dynamics at lower energies (top RHIC energies) and in smaller collision systems such as Ne-Ne and O-O, which recently collided at the LHC. In all these cases, the applicability of a hydrodynamic description to a heavy particle is more uncertain, making it an interesting object of study, as we will clarify in the next sections.
\section{Fluid dynamics for heavy quarks}
The fluid-dynamic equations to solve are mainly given by the system of equations
\begin{gather}
    \nabla_\mu T^{\mu\nu} = 0\,,\quad
    \nabla_\mu N^\mu = 0\,,
\end{gather}
which expresses the conservation of the energy-momentum tensor $T^{\mu\nu}$ and of an additional conserved current $N^\mu$. The latter is associated with conserving the number of charm-anticharm pairs~\cite{Capellino:2022nvf}.
The Landau frame is chosen such that $T^{\mu\nu}$ and $N^\mu$ can be decomposed as
\begin{gather}
    T^{\mu \nu} = (\epsilon+p)u^\mu u^\nu +\Delta^{\mu\nu}(p+\Pi)+\pi^{\mu\nu}\,,\quad
    N^\mu = n u^\mu + \nu^\mu\,,
\end{gather}
where $\epsilon$, $p$, $u^\mu$, $\Pi$ and $\pi^{\mu\nu}$ are the energy density, thermodynamic pressure, fluid four-velocity, bulk viscous pressure, and shear-stress tensor of the fluid, respectively. The charm-quark fields are the heavy-quark density $n$ and the diffusion current $\nu^\mu$. The local temperature $T$ and the charm chemical potential to temperature ratio $\alpha$ are determined by the Landau matching conditions.
We assume that the energy density is independent of the heavy-quark contribution, such that any energy density dependence on $\alpha$ is negligible. 
The equilibrium heavy-quark density is taken to be one of the hadron-resonance gas, including all measured charm states (HRGc)~\cite{Capellino:2023cxe}.
The equations of motion for each of the dissipative currents are solved in a second-order hydrodynamic formalism. We report here the equation for the heavy-quark diffusion current,
\begin{align}
     \tau_n \Delta^\alpha_\beta u^\mu \nabla_\mu \nu^\beta + \nu^\alpha + \kappa_n \Delta^{\alpha\beta} \partial_\beta \alpha &=0\label{eq:nueq}
     \,,
\end{align}
where we introduced the heavy-quark diffusion coefficient $\kappa_n$ and its corresponding relaxation time $\tau_n$. We remark that $\kappa_n$ and $\tau_n$ are proportional to the heavy-quark spatial diffusion coefficient $D_s$~\cite{Capellino:2022nvf}. 
The equations are solved in Bjorken coordinates, assuming boost and azimuthal rotation invariance.

\label{sec-1}
\section{Results}
The success of the fluid-dynamic description for charm quarks at low transverse momenta at LHC energies is supported by the results reported in Ref.~\cite{Capellino:2023cxe}. In this contribution, however, we want to explore new directions that test the very applicability of the fluid-dynamic description for heavy quarks. 
We are going to probe the dynamics of charm quarks in Au-Au collisions at top RHIC energies and to O-O and Ne-Ne collisions at LHC energies. In both cases, we face similar challenges: the charm production cross section is smaller in these collision systems than in Pb-Pb, resulting in a lower charm abundance. Furthermore, since the temperatures reached within these collisions are less than in the Pb-Pb case, the relaxation times will be larger. The O-O and Ne-Ne collisions present additional features coming from the nuclear deformation of the colliding nuclei, which highly impact the flow coefficients. However, in this study, we focus on transverse momentum spectra and integrated yields only, which are less affected by this.

\label{sec-2}

\subsection{Charm in Au-Au collisions at RHIC}
\begin{figure}
    \centering
    \includegraphics[width=0.8\linewidth]{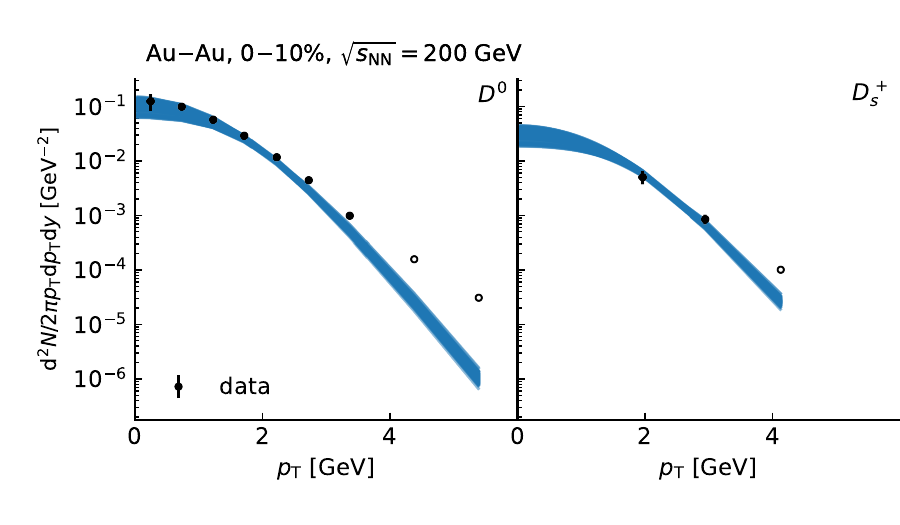}
    \caption{Transverse momentum distribution of $\mathrm{D^0}$ and $\mathrm{D_s^+}$ in comparison with the experimental measurements \cite{STAR:2018zdy,STAR:2021tte} in Au-Au collisions at 200 GeV in the 0-10$\%$ centrality class.}
    \label{fig:auau_200}
\end{figure}

The richness of experimental measurements for open charm hadrons in Au-Au collisions at 200 GeV at RHIC makes this an excellent platform to test our fluid-dynamic framework for charm and gain further information on the charm spatial diffusion coefficient. In Fig.~\ref{fig:auau_200} we report the results for the transverse momentum distribution of $\mathrm{D_0}$ and $\mathrm{D_s^+}$ in central collisions in comparison with the experimental measurements from the STAR collaboration~\cite{STAR:2018zdy,STAR:2021tte}. The colored bands are our model results after performing a Bayesian analysis with available RHIC data to extract the spatial diffusion coefficient~\cite{rossana}. Remarkably, despite only 2-3 charm-anticharm pairs being produced at these energies per unit rapidity, the fluid dynamic description seems to be applicable and describes the experimental data up to $p_{\rm T}\sim3$~GeV.
\subsection{Charm in O-O and Ne-Ne collisions at the LHC}
\begin{figure}
    \centering
    \includesvg[width=0.49\linewidth]{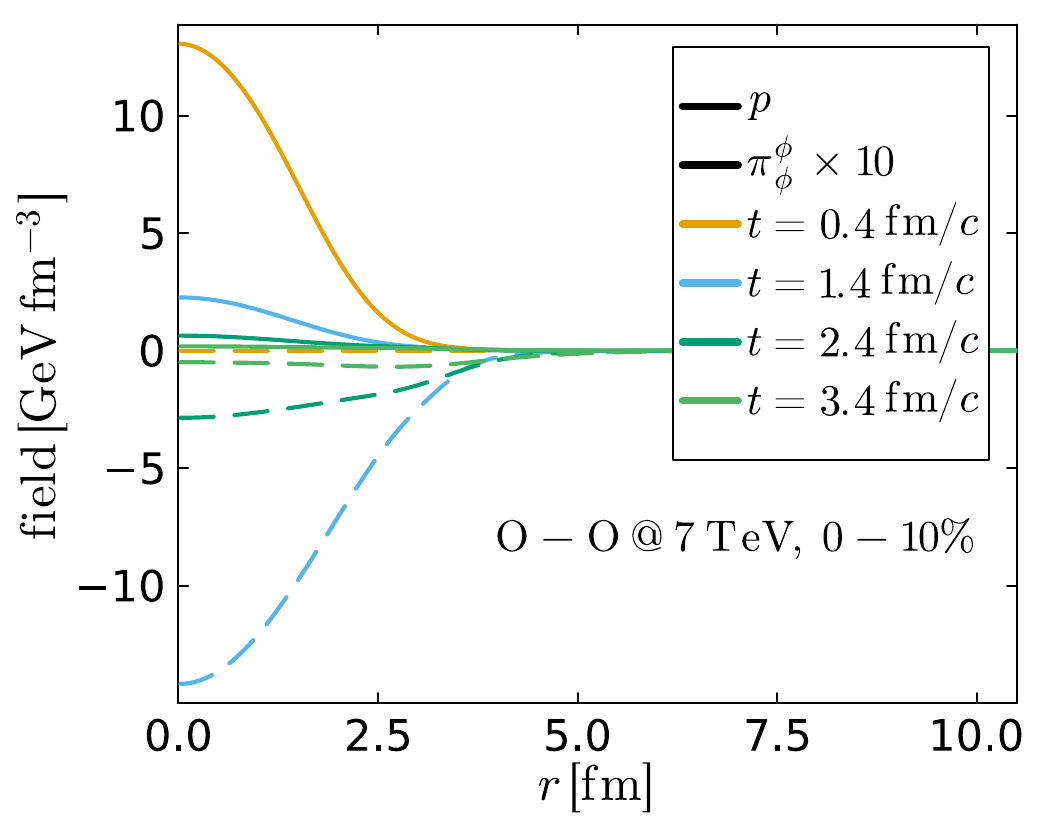}
    \includesvg[width=0.49\linewidth]{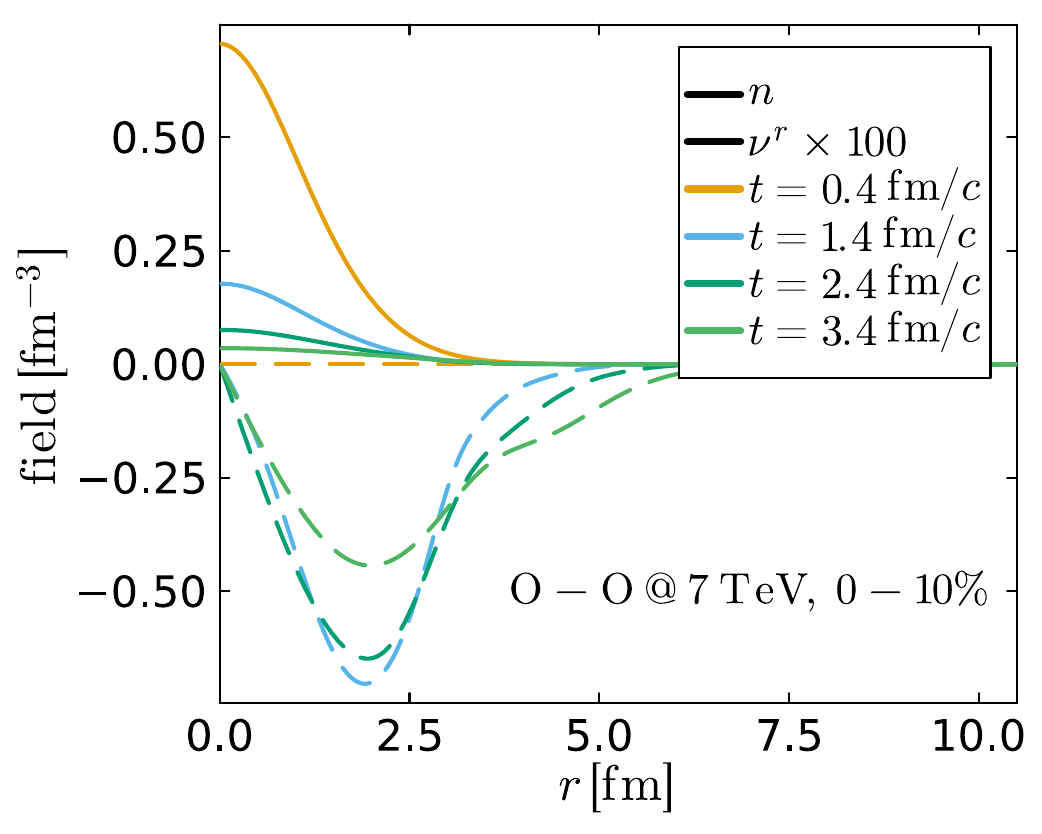}
    \caption{Evolution of hydrodynamic fields in O-O collisions at 7 TeV in the 0-10$\%$ centrality class. Left panel: equilibrium pressure $p$ and shear-stress tensor component $\pi^\phi_\phi$ as a function of radius at different Bjorken times. Right panel: charm equilibrium density $n$ and diffusion current component $\nu^r$.}
    \label{fig:OO}
\end{figure}
In O-O and Ne-Ne collisions, there is an additional layer of complexity. In fact, it is currently still unclear whether fluid dynamics is applicable at all for this type of collision, and if a QGP is produced in the sense of a thermalized state of deconfined quarks and gluons. Therefore, we first must check that, assuming a QGP is produced, the fluid-dynamic description of the bulk is stable. In the left panel in Fig.~\ref{fig:OO} we plot the equilibrium pressure $p$ and the shear-stress tensor component $\pi^\phi_\phi$ as a function of radius at different Bjorken times in O-O central collisions (similar results are observed for Ne-Ne). The equilibrium pressure is always found to be larger than the out-of-equilibrium component, allowing for a positive total pressure in the entire spatial domain, and the shear stress correction correctly relaxes towards equilibrium. 
Applying the same concept to the heavy quark dynamics, in the right panel, we look at the charm density $n$ and the diffusion current component $\nu^r$. The condition $|\nu^r|\ll n$ always holds in the region occupied by the fireball, and the diffusion current relaxes towards equilibrium. This paves the way for a hydrodynamic treatment of charm dynamics in smaller collision systems. The validity of such an approach, however, can only be confirmed by experimental measurements.
\section{Conclusions}
We explored new directions regarding the applicability of fluid dynamics to study heavy-quark diffusion in the QGP. The fluid-dynamic description seems to be well-defined and stable in the investigated scenarios, namely charm diffusion at RHIC top energies and in O-O and Ne-Ne collisions at the LHC. We look forward to new experimental data from the ongoing Run3 at the LHC, which can confirm or disprove the extension of the hydrodynamic description to these limit-case scenarios. Furthermore, we are going to extend this work to beauty quarks to investigate whether fluid dynamics is applicable to a heavier particle.
%
\bibliography{bibliography.bib} 

\begin{thebibliography}{9}

\bibitem{ALICE:2020iug}
S.~Acharya et~al. (ALICE), {Transverse-momentum and event-shape dependence of
  D-meson flow harmonics in Pb\textendash{}Pb collisions at $\sqrt {s_{NN}}$ =
  5.02 TeV}, Phys. Lett. B \textbf{813}, 136054 (2021), \texttt{2005.11131}.
  \doiwoc{10.1016/j.physletb.2020.136054}

\bibitem{ALICE:2020pvw}
S.~Acharya et~al. (ALICE), {J/$\psi$ elliptic and triangular flow in Pb-Pb
  collisions at $\sqrt{s_{\rm NN}}$ = 5.02 TeV}, JHEP \textbf{10}, 141 (2020),
  \texttt{2005.14518}. \doiwoc{10.1007/JHEP10(2020)141}

\bibitem{Andronic:2021erx}
A.~Andronic, P.~Braun-Munzinger, M.K. K\"ohler, A.~Mazeliauskas, K.~Redlich,
  J.~Stachel, V.~Vislavicius, {The multiple-charm hierarchy in the statistical
  hadronization model}, JHEP \textbf{07}, 035 (2021), \texttt{2104.12754}.
  \doiwoc{10.1007/JHEP07(2021)035}

\bibitem{Altenkort:2023oms}
L.~Altenkort, O.~Kaczmarek, R.~Larsen, S.~Mukherjee, P.~Petreczky, H.T. Shu,
  S.~Stendebach (HotQCD), {Heavy Quark Diffusion from 2+1 Flavor Lattice QCD
  with 320~MeV Pion Mass}, Phys. Rev. Lett. \textbf{130}, 231902 (2023),
  \texttt{2302.08501}. \doiwoc{10.1103/PhysRevLett.130.231902}

\bibitem{Capellino:2022nvf}
F.~Capellino, A.~Beraudo, A.~Dubla, S.~Floerchinger, S.~Masciocchi,
  J.~Pawlowski, I.~Selyuzhenkov, {Fluid-dynamic approach to heavy-quark
  diffusion in the quark-gluon plasma}, Phys. Rev. D \textbf{106}, 034021
  (2022), \texttt{2205.07692}. \doiwoc{10.1103/PhysRevD.106.034021}

\bibitem{Capellino:2023cxe}
F.~Capellino, A.~Dubla, S.~Floerchinger, E.~Grossi, A.~Kirchner, S.~Masciocchi,
  {Fluid dynamics of charm quarks in the quark-gluon plasma}, Phys. Rev. D
  \textbf{108}, 116011 (2023), \texttt{2307.14449}.
  \doiwoc{10.1103/PhysRevD.108.116011}

\bibitem{STAR:2018zdy}
J.~Adam et~al. (STAR), {Centrality and transverse momentum dependence of
  $D^0$-meson production at mid-rapidity in Au+Au collisions at ${\sqrt{s_{\rm
  NN}} = \rm{200\,GeV}}$}, Phys. Rev. C \textbf{99}, 034908 (2019),
  \texttt{1812.10224}. \doiwoc{10.1103/PhysRevC.99.034908}

\bibitem{STAR:2021tte}
J.~Adam et~al. (STAR), {Observation of $D_{s}^{\pm}/D^0$ enhancement in Au+Au
  collisions at $\sqrt{s_{_{NN}}}$ = 200 GeV}, Phys. Rev. Lett. \textbf{127},
  092301 (2021), \texttt{2101.11793}. \doiwoc{10.1103/PhysRevLett.127.092301}

\bibitem{rossana}
R.~Facen, Master's thesis, PI Heidelberg (2024), available
  \href{https://www.physi.uni-heidelberg.de/Publications/RossanaFacen_MasterThesis.pdf}{here}

\end{thebibliography}

\end{document}